\documentclass[preprint,showpacs,amsmath,amssymb]{revtex4}
\usepackage{mathrsfs}

\usepackage{graphicx,color}
\usepackage{dcolumn}
\usepackage{bm}
\usepackage{CJK}

\begin{document}
\begin{CJK*}{GBK}{song}
%-------------------------------------------------------------------------------------------------------

\title{Nuclear effective charge factor originating from covariant density functional theory}

\author{Z. M. Niu$^1$}\email{zmniu@ahu.edu.cn}
\author{Q. Liu$^1$}\email{quanliu@ahu.edu.cn}
\author{Y. F. Niu$^2$}%\email{nyfster@gmail.com}
\author{W. H. Long$^3$}%\email{longwh@lzu.edu.cn}
\author{J. Y. Guo$^1$}\email{jianyou@ahu.edu.cn}

\affiliation{$^1$School of Physics and Material Science, Anhui University,
             Hefei 230039, China}
\affiliation{$^2$Institute of Fluid Physics, China Academy of Engineering
             Physics, Mianyang 621900, China}
\affiliation{$^3$School of Nuclear Science and Technology, Lanzhou
             University, Lanzhou 730000, China}

\date{\today}

%-------------------------------------------------------------------------------------------------------
\begin{abstract}
Guiding by the relativistic local density approximation, we explore a
phenomenological formula for the coupling strength of Coulomb field to
take into account the Coulomb exchange term effectively in the
relativistic Hartree approximation. Its validity in finite nuclei is
examined by comparing with the exact treatment of the Coulomb exchange
energies in the relativistic Hartree-Fock-Bogoliubov approach. It is
found that the exact Coulomb exchange energies can be reproduced by
employing the phenomenological formula with the relative deviations less
than 1\% for semi-magic Ca, Ni, Sn, and Pb isotopes. Furthermore, we
check the applicability of the phenomenological formula for the
effective interactions in the relativistic Hartree approach by
investigating the binding energy differences of mirror nuclei.
\end{abstract}

 \maketitle

%-------------------------------------------------------------------------------------------------------
%The study of exotic nuclei far from the $\beta$-stability line has become
%an active field of research, as lots of RIB facilities are operating and
%being upgraded~\cite{Tanihata1985PRL, Meng1996PRL, Ozawa2000PRL,
%Zilges2005PPNP, Otsuka2010PRL}.
As one of the most significant building blocks in nuclei, the Coulomb
interaction between protons in nuclei plays an important role in
understanding many nuclear phenomena, such as the Coulomb displacement
energies, isospin mixing, proton emission, and fission barriers. It
therefore requires an efficient and precise treatment of the Coulomb
effects for the reliable description of nuclear structure
properties~\cite{Auerbach1983PRp, Meng2006PPNP, Vretenar2005PRp}.

%In the nuclear mean-field theory, there are direct (Hartree) and exchange
%(Fock) terms for the Coulomb interaction. However, the exchange term is
%very cumbersome and time-consuming in the practical calculations due to
%the non-locality of the corresponding mean field. Therefore, the Coulomb
%exchange term is usually considered within a local scheme by using the
%so-called Slater approximation for the non-relativistic self-consistent
%Hartree-Fock calculations in the Skyrme~\cite{Schnaider1974PLB,
%Skalski2001PRC, Bloas2011PRC} or Gogny~\cite{Anguiano2001NPA} approaches.

During the past years, the density functional theory (DFT) with the
Lorentz symmetry, namely, the covariant density functional theory (CDFT),
has received wide attention due to many successes achieved in describing
lots of nuclear phenomena~\cite{Meng2006PPNP, Vretenar2005PRp} as well as
its successful applications in the astrophysics~\cite{Sun2008PRC,
Niu2009PRC, Niu2011PRC, Niu2012arXiv, Niu2011IJMPE, Meng2011SCSG,
Zhang2012APS}.
%There exists a number of attractive features in the RMF theory: naturally
%includes the nucleon spin degree of freedom, and the resulting nuclear
%spin-orbit potential automatically emerges with the empirical strength,
%thus producing a good agreement with the experimental single-nucleon
%spectrum~\cite{Ring1996PPNP, Liang2011PRC}; gives more naturally the
%origin of the pseudospin symmetry as a relativistic symmetry and the spin
%symmetry in the anti-nucleon spectrum~\cite{Ginocchio1997PRL, Meng1998PRC,
%Meng1999PRC, Zhou2003PRL}; better reproduces the isotopic shifts in the Pb
%region~\cite{Sharma1993PLB}.
Specifically, there exist two widely used approaches in the CDFT
framework: the relativistic Hartree (RH)~\cite{Walecka1974AP,
Meng2006PPNP, Vretenar2005PRp} and relativistic Hartree-Fock (RHF)
approaches~\cite{Bouyssy1985PRL, Long2006PLB}. The former one is usually
known as the relativistic mean field (RMF) model.

Compared with simple direct (Hartree) term, the Coulomb exchange (Fock)
term is very cumbersome and time-consuming in the practical calculations
due to the non-locality commonly existing in the Fock mean field. To keep
the consistent with the approach itself and retain the simplicity of the
theory, the non-local Coulomb exchange term is usually neglected in the RH
framework. Its effects, in principle, can be taken into account partially
by the parametrization of the effective coupling strengths of the model.
In the recent years, the RHF approach has been well developed with
substantial improvements in the self-consistent descriptions of nuclear
shell structure and the evolution, the restoration of relativistic
symmetry, exotic nuclei, the low-energy excitation mode,
etc.~\cite{Long2006PLB, Long2007PRC, Long2008EL, Long2009PLB, Long2010PRC,
Ebran2011PRC, Liang2008PRL, Liang2012PRC}. Additionally it is also found
from the RHF approach that the prescription of neglecting the Coulomb
exchange term in RH approach is not always valid. One example is the
isospin symmetry-breaking corrections to the superallowed $\beta$ decays,
which are crucial for testing the unitarity of the
Cabibbo-Kobayashi-Maskawa matrix~\cite{Liang2009PRC}. It is therefore
desirable for the RH approach to get the Coulomb exchange effects involved
efficiently with satisfied accuracy.

In the non-relativistic framework, the Coulomb exchange term is usually
evaluated within the local density approximation (LDA), which is the
well-known Slater approximation~\cite{Slater1951PR}. The validity of this
approximation has been investigated in the Skyrme~\cite{Schnaider1974PLB,
Skalski2001PRC, Bloas2011PRC} or Gogny~\cite{Anguiano2001NPA} approaches.
The relativistic local density approximation (RLDA) for the Coulomb
exchange functional in nuclear systems, which is the Slater approximation
with relativistic corrections, has been developed
recently~\cite{Gu2012arXiv}. It is found that the relativistic corrections
can remarkably improve the description of the exact Coulomb exchange
energies in the relativistic Hartree-Fock-Bogoliubov (RHFB) approach and
the relative deviations are less than $5\%$ for semi-magic isotopes.
However, there still exist some systematic deviations between the
self-consistent RLDA calculations and the exact values for heavy nuclei,
such as Sn and Pb isotopes.

%Recently, exact RHF approach has been developed and good description for
%finite nuclei and nuclear matter has been achieved with a number of
%adjustable parameters comparable to that of the RH
%approach~\cite{Long2006PLB, Long2010PRC, Ebran2011PRC}. It has been found
%the inclusion of exchange terms in RHF approach can improve the
%description of the nuclear shell structure~\cite{Long2006PLB, Long2007PRC}
%and its evolution~\cite{Long2008EL, Long2009PLB}, achieve excellent
%agreement with data on the Gamow-Teller (GTR) and spin-dipole resonances
%(SDR)~\cite{Liang2008PRL, Liang2012PRC}.

Within the RH approach, the meson-nucleon coupling strengths are
parameterized to obtain appropriate quantitative description of nuclear
structure properties and consequently part of the effects beyond Hartree
and no-sea approximation can be taken into account effectively. Following
this spirit, we will explore a phenomenological formula for the coupling
strength of Coulomb field to include the Coulomb exchange term effectively
in the RH approximation. In this work, we first explore the form of this
phenomenological formula guiding by the RLDA, and then examine its
validity by comparing with the exact treatment of the Coulomb exchange
term in the RHFB approach~\cite{Long2010PRC}. Furthermore, we employ this
formula to investigate the binding energy differences of mirror nuclei and
check its validity in the RH approach, with the effective interactions
NL3~\cite{Lalazissis1997PRC}, PKDD~\cite{Long2004PRC}, and
DD-ME2~\cite{Lalazissis2005PRC}.

In the RHF framework, the Coulomb energy $E_{\rm C}$ for the nuclear
system with time-reversal symmetry consists of the direct term $E_{\rm
Cdir}$ and the exchange one $E_{\rm Cex}$,
\begin{eqnarray}
  E_{\rm Cdir}=\frac{e^2}{2}\int\int d^3r d^3r'
               \frac{\rho_p(\boldsymbol{r})\rho_p(\boldsymbol{r'})}
               {|\boldsymbol{r}-\boldsymbol{r'}|},
\end{eqnarray}
\begin{eqnarray}
  E_{\rm Cex}=-\frac{e^2}{2}&&\sum_{ij}^p v_i^2 v_j^2 \int\int d^3r d^3r'
                 \frac{\cos(|\varepsilon_i-\varepsilon_j||\boldsymbol{r}-\boldsymbol{r'}|)}
                 {|\boldsymbol{r}-\boldsymbol{r'}|}\nonumber\\
              & &\times~~\bar{\psi}_i(\boldsymbol{r}) \gamma^\mu \psi_j(\boldsymbol{r})
                        \bar{\psi}_j(\boldsymbol{r'}) \gamma_\mu \psi_i(\boldsymbol{r'}),
\end{eqnarray}
where $\rho_p(\boldsymbol{r})$, $\varepsilon_i$, $v_i^2$ and $\psi_i$
denote the proton density, the single-particle energy, occupation
probability, and wave function, respectively~\cite{Gu2012arXiv}. By
introducing the effective charge factor
\begin{equation}\label{Eq:ExactChFac}
   \eta=\sqrt{1+\frac{E_{\rm Cex}}{E_{\rm Cdir}}},
\end{equation}
the Coulomb energy $E_{\rm C}$ can be expressed as
\begin{equation}
   E_{\rm C}
  =\eta^2 E_{\rm Cdir}
  =\frac{\left(\eta e\right)^2}{2}\int\int d^3r d^3r'
               \frac{\rho_p(\boldsymbol{r})\rho_p(\boldsymbol{r'})}
               {|\boldsymbol{r}-\boldsymbol{r'}|}.
\end{equation}
Then the $E_{\rm Cex}$ can be taken into account effectively in the
Hartree approximation by merely changing $e$ to $\eta e$.

As referred to the RLDA, another effective way to include the exchange
contributions of Coulomb field into the RH approach~\cite{Gu2012arXiv}, a
phenomenological formula is firstly developed for the effective charge
factor $\eta$. In the RLDA, the $E_{\rm Cex}$ can be expanded with respect
to the $\beta=(3\pi^2 \rho_p)^{1/3}/M$, where $M$ is the proton mass. Up
to the second order of $\beta$, the Coulomb exchange energy in the RLDA
reads as
\begin{equation}\label{Eq:ECexRLDA}
  E_{\rm Cex}^{\rm RLDA}
  = -\frac34\left(\frac3\pi\right)^{1/3}e^2\int d^3r\rho_p^{4/3}
    \left(1-\frac23 \beta^2\right).
\end{equation}
Empirically the proton density can be approximated as
\begin{equation}\label{Eq:PheProtDen}
  \rho_p(r)=\left\{\begin{array}{cc}
                0~,             &~~r> R\\
                Z/(4\pi R^3/3)~,&~~r\leqslant R
                \end{array}~~,\right.
\end{equation}
with nuclear radius $R=r_0 A^{1/3}$ and then
\begin{equation}
   \frac{E_{\rm Cex}^{\rm RLDA}}{E_{\rm Cdir}}
  =-\frac54\left(\frac{3}{2\pi}\right)^{2/3}Z^{-2/3}
   +\frac{15}{8M^2r_0^2}A^{-2/3}.
\end{equation}

From the calculations of the RHFB approach with the effective interaction
PKA1~\cite{Long2007PRC}, it is found that the exact $E_{\rm Cex}$ is much
smaller than $E_{\rm Cdir}$. Taking the $^{132}$Sn as an example, the
absolute ratio between $E_{\rm Cex}$ and $E_{\rm Cdir}$ is about $0.05$.
Therefore, one can expand the effective charge factor $\eta$ as the powers
of $E_{\rm Cex}/E_{\rm Cdir}$, i.e.,
\begin{equation}
   \eta
  =\sqrt{1+\frac{E_{\rm Cex}}{E_{\rm Cdir}}}
  =1 + \frac12\frac{E_{\rm Cex}}{E_{\rm Cdir}}
     - \frac18\left(\frac{E_{\rm Cex}}{E_{\rm Cdir}}\right)^2
     + \cdot\cdot\cdot~~,
\end{equation}
Up to the linear order of $(E_{\rm Cex}/E_{\rm Cdir})$, the effective
charge factor determined by the RLDA can be written as
\begin{equation}\label{Eq:PheChFac}
   \eta
  =1 -\frac58\left(\frac{3}{2\pi}\right)^{2/3}Z^{-2/3}
   +\frac{15}{16M^2r_0^2}A^{-2/3}.
\end{equation}
Inspired by the above formalism, we employ the expression
\begin{equation}\label{Eq:EffChFac}
   \eta(Z,A) = 1 - aZ^b + cA^d
\end{equation}
to parameterize the effective charge factor, as referred to the exact
calculations of RHFB.

%In this work, the RHFB calculation adopts the effective interaction
%PKA1~\cite{Long2007PRC} and D1S~\cite{Berger1991CPC} for the particle-hole
%and particle-particle channels respectively.
%The RHFB equations are solved on a Dirac Woods-Saxon basis
%within a spherical box of radius $R_{\rm max} = 20$ fm, and the numbers of
%positive and negative energy levels for each state are fixed to $N_F = 28$
%and $N_D = 20$, respectively.
%For simplicity, the Coulomb exchange energies calculated in the standard
%RHFB approach are labeled with the ``exact". The corresponding results
%calculated with the effective charge factor in Eq.~(\ref{Eq:EffChFac}) are
%label with the ``$\eta(Z, A)$". For comparison, the results of
%nonrelativistic and relativistic LDA taken from Ref.~\cite{Gu2012arXiv}
%are also presented, which are labeled with the ``NRLDA" and ``RLDA",
%respectively.

\begin{figure}[h]
  \includegraphics[width=7cm]{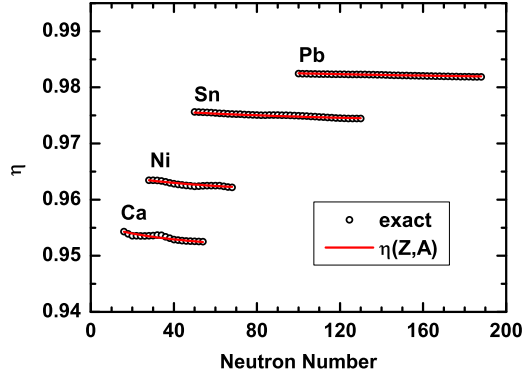}\\
  \caption{(Color online) The effective charge factors for the semi-magic Ca,
  Ni, Sn, and Pb isotopes calculated by the RHFB approach with PKA1. The
  fitted values with Eq.~(\ref{Eq:EffChFac}) are shown by solid lines.}
  \label{fig1}
\end{figure}

In this work, the effective interactions PKA1~\cite{Long2007PRC} and
D1S~\cite{Berger1991CPC} are utilized in the particle-hole and
particle-particle channels for the RHFB calculations, respectively. With
the calculated $E_{\rm Cex}$ and $E_{\rm Cdir}$ in the RHFB approach, the
exact effective charge factors can be obtained for each nucleus using
Eq.~(\ref{Eq:ExactChFac}). The exact effective charge factors for
traditional semi-magic Ca, Ni, Sn, and Pb isotopes are shown in
Fig.~\ref{fig1} by the open circles. By fitting to these exact effective
charge factors with Eq.~(\ref{Eq:EffChFac}), the parameters are determined
as $a=0.366958, b=-0.645775, c=0.030379, d=-0.398341$ and the
corresponding effective charge factors are shown in Fig.~\ref{fig1} by the
solid lines. It is clear that the effective charge factor is sensitive to
the proton number and increases as the proton number increases. In
addition, there also exists a weak isospin dependence along an isotopic
chain, which slightly decreases as the neutron number increases in
general. With the phenomenological formula in Eq.~(\ref{Eq:EffChFac}), the
fitted effective factors are in excellent agreement with the exact values.
Comparing Eq.~(\ref{Eq:EffChFac}) with Eq.~(\ref{Eq:PheChFac}), it is
found that the value of the parameter $b$ is close to $-2/3$, while the
parameter $d$ shows a remarkable deviation from $-2/3$. In fact, the
deviation mainly originates from the higher-order dependence on the proton
density, i.e., the $\beta^2$ term in Eq.~(\ref{Eq:ECexRLDA}), introduced
by the relativistic correction to $E_{\rm Cex}$ in the LDA. For the proton
density, we adopt a phenomenological formula in Eq.~(\ref{Eq:PheProtDen}),
which is too simple to well describe the real nuclear system. Therefore,
by taking the coefficients in Eq.~(\ref{Eq:PheChFac}) as free parameters,
the deviations from real nuclear systems can be taken into account
effectively.

%\begin{table}
%\begin{center}
%\caption{The parameters $a, b, c, d$ in Eq.~(\ref{Eq:EffChFac}).}\label{Tab:EffChFac}
%\begin{tabular}{ccccccc}
%\hline\hline
%a          &~~    &b          &~~   &c         &~~   &d\\
%\hline
%0.366958   &~~    &-0.645775  &~~   &0.030379  &~~   &-0.398341\\
%\hline\hline
%\end{tabular}
%\end{center}
%\end{table}

\begin{figure}[h]
  \includegraphics[width=7cm]{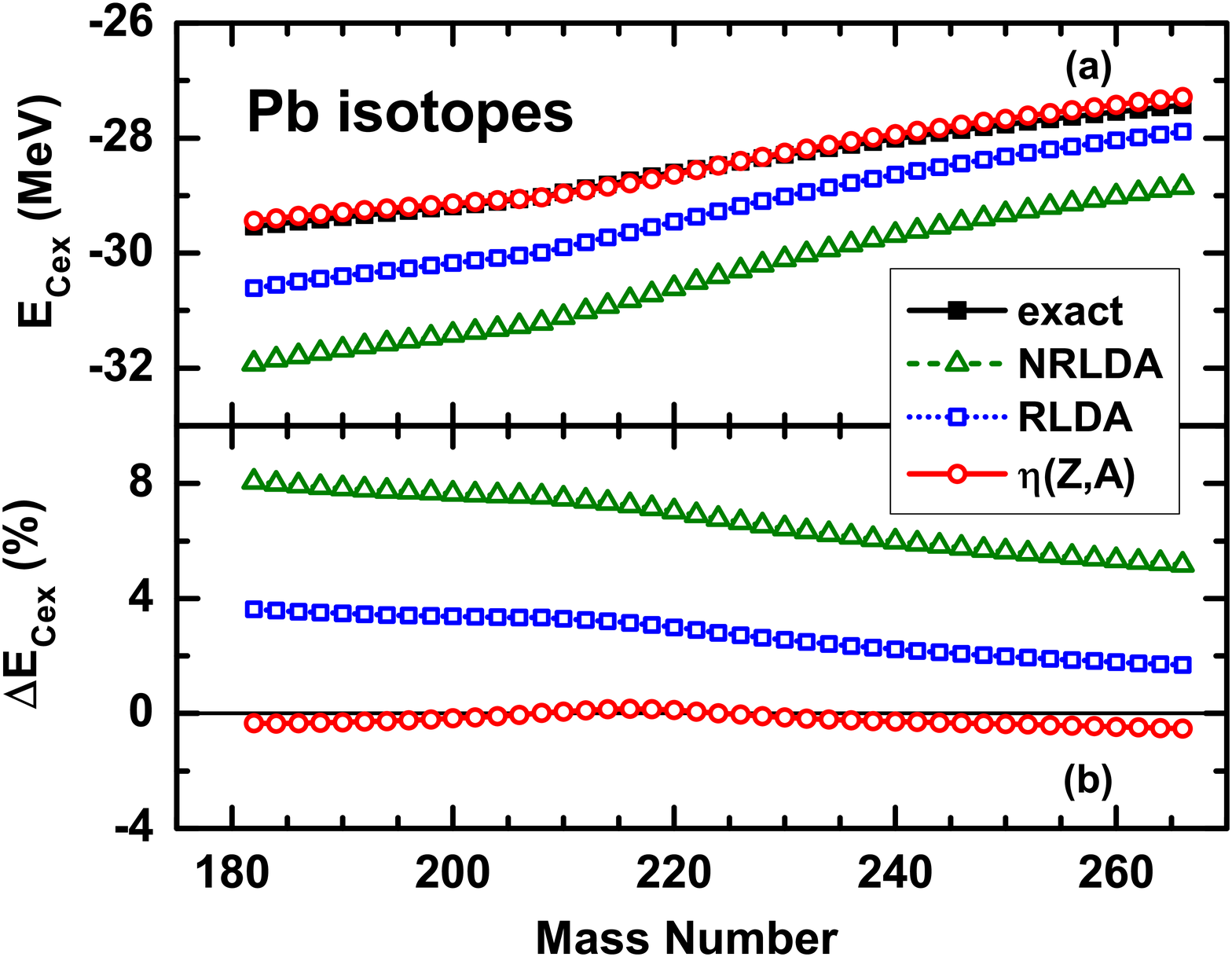}
  \caption{(Color online) Coulomb exchange energies and the corresponding
  relative deviations from the exact results in Pb isotopes obtained from
  phenomenological effective charge factor (open circles). For comparison,
  the exact results from the RHFB approach with PKA1 and other results
  obtained with the NRLDA and RLDA are shown by the filled squares, open
  triangles, and open squares, respectively.}
  \label{fig2}
\end{figure}

In order to exclude the effects due to the self-consistency, one-step
calculations have been performed to investigate the effects of the LDA on
Coulomb exchange energies in Ref.~\cite{Gu2012arXiv}, i.e., $E_{\rm
Cex}^{\rm NRLDA}$ and $E_{\rm Cex}^{\rm RLDA}$, are respectively obtained
from the nonrelativistic local density approximation (NRLDA) and RLDA with
the proton density $\rho_p(r)$ given by the self-consistent RHFB
calculations. In terms of the phenomenological effective charge factor,
the Coulomb exchange contributions can be approximated as
\begin{equation}
   E_{\rm Cex}^{\eta(Z,A)} = \left(1-\frac1{\eta^2}\right)E_{\rm C}.
\end{equation}
The relative deviations of the approximate Coulomb exchange energies
$E_{\rm Cex}^{\rm approx}$ from the exact RHFB results are defined as
\begin{equation}\label{Eq:RelDevEex}
   \Delta E_{\rm Cex}
   =\frac{E_{\rm Cex}^{\rm approx}-E_{\rm Cex}^{\rm exact}}{E_{\rm Cex}^{\rm exact}}.
\end{equation}
By taking the even-even Pb isotopes from proton drip line to neutron drip
line as examples, the calculated Coulomb exchange energies $E_{\rm
Cex}^{\textrm{exact}}$, $E_{\rm Cex}^{\textrm{NRLDA}}$, $E_{\rm
Cex}^{\textrm{RLDA}}$, and $E_{\rm Cex}^{\eta(Z,A)}$ are shown as a
function of mass number in the panel (a) of Fig.~\ref{fig2}. The
corresponding relative deviations $\Delta E_{\rm Cex}$ defined in
Eq.~(\ref{Eq:RelDevEex}) are shown in the panel (b) of Fig.~\ref{fig2}.

It is found that the magnitudes of the Coulomb exchange energies $E_{\rm
Cex}$ generally decrease with increasing mass number. For each nucleus,
the magnitude of $E_{\rm Cex}$ is overestimated by the NRLDA, and
substantially improved when the relativistic corrections are taken into
account~\cite{Gu2012arXiv}. However, the calculations with the RLDA still
overestimated the exact $E_{\rm Cex}$ by about $1$ MeV, especially for the
neutron-deficient nuclei. Further improvement on the description of the
exact $E_{\rm Cex}$ is achieved for the calculations with the
phenomenological effective charge factors in Eq.~(\ref{Eq:EffChFac}). From
the panel (b) of Fig.~\ref{fig2}, it is found that the relative deviations
for the calculations with the NRLDA and the RLDA are respectively
$5.2\%\sim 8.1\%$ and $1.7\%\sim 3.6\%$, while the relative deviation for
the calculations with the phenomenological effective charge factor is
smaller than $0.5\%$. Moreover, the systematic deviations from the exact
values are eliminated. This shows that the contributions beyond the RLDA
can be effectively included by using the phenomenological effective charge
factors in Eq.~(\ref{Eq:EffChFac}).

\begin{figure}[h]
  \includegraphics[width=8cm]{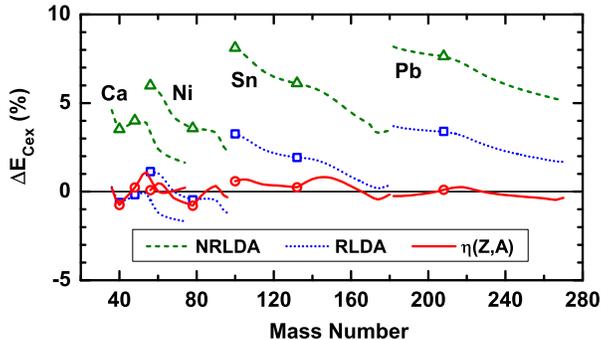}
  \caption{(Color online) Relative deviations of the Coulomb exchange energies
  by the self-consistent calculations with the effective charge factor (solid
  lines), NRLDA (dashed lines) and RLDA (dotted lines) for Ca, Ni, Sn,
  and Pb isotopes. The traditional doubly magic nuclei are denoted as open
  symbols.}
  \label{fig3}
\end{figure}

In order to investigate the effects due to the self-consistency, the
self-consistent calculations with phenomenological effective charge
factors have been performed as well. The systematics of calculated results
for the semi-magic Ca, Ni, Sn, and Pb isotopes from proton drip line to
neutron drip line are shown in Fig.~\ref{fig3}, where the traditional
doubly magic nuclei are marked by the open symbols. Comparing with the
exact Coulomb exchange energies from the RHFB calculations, it is found
that the relative deviations $\Delta E_{\rm Cex}$ for the calculations
with the phenomenological effective charge factor are less than $1\%$ for
the selected semi-magic isotopes, while the results with the NRLDA show
remarkable systematic deviations and the maximum deviation even exceeds
$8\%$. For the Ca and Ni isotopes, the self-consistent calculations with
the phenomenological effective charge factor and the RLDA show a similar
accuracy. As the proton number increases, the RLDA calculation
systematically overestimates the magnitude of Coulomb exchange energies
for the Sn and Pb isotopes, leading to a large relative deviation, while
the relative deviation for the calculation with phenomenological effective
charge factor is still within $1\%$. Therefore, the phenomenological
effective charge factor shown in Eq.~(\ref{Eq:EffChFac}) for the Coulomb
exchange term in nuclear CDFT is more robust than the NRLDA and RLDA. In
particular, the systematic deviations from the exact Coulomb exchange
energies for the NRLDA and RLDA are eliminated not only for the light
nuclei, but also for the heavy nuclei.

\begin{figure}[h]
  \includegraphics[width=7cm]{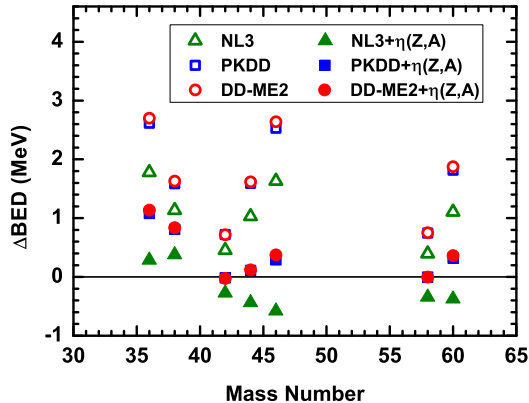}
  \caption{(Color online) The discrepancy between the BED of mirror nuclei
  calculated by RHB approach with the parameter sets, i.e.,
  NL3 (triangles), PKDD (squares), or DD-ME2 (circles), and those calculated
  by RHFB approach with PKA1. The filled and open symbols denote the
  results with and without the phenomenological effective charge
  factors, respectively.}
  \label{fig4}
\end{figure}

Since the binding energy differences (BED) of mirror nuclei are sensitive
to the nuclear Coulomb field, the exchange term of Coulomb field
inevitably plays an important role in understanding the BED of mirror
nuclei. In Fig.~\ref{fig4}, we display the discrepancy between the BED of
mirror nuclei calculated by RHB approach with the parameter sets, i.e.,
NL3~\cite{Lalazissis1997PRC}, PKDD~\cite{Long2004PRC}, or
DD-ME2~\cite{Lalazissis2005PRC}, and those calculated by RHFB approach
with PKA1. The results with and without the phenomenological effective
charge factors are shown by filled and open symbols, respectively. It is
clear that the BED of mirror nuclei calculated with the RH parameter sets
are systematically larger than those calculated by the RHFB approach with
PKA1. When the phenomenological effective charge factors are employed in
the RHB calculations, the discrepancy is remarkably reduced and the
systematic deviation is eliminated. This improvement shows that the
phenomenological effective charge factors obtained in this work are still
applicable for these RH parameters.

In summary, guiding by the RLDA, we explore a phenomenological formula for
the coupling strength of Coulomb field to take into account the Coulomb
exchange term effectively in the RH approximation. Comparing with the
NRLDA and RLDA, the description of exact Coulomb exchange energies in the
RHFB calculations is remarkably improved with the phenomenological
effective charge factors. In particular, the systematic deviations of the
NRLDA and RLDA calculations from the exact Coulomb exchange energies are
eliminated not only for the light nuclei, but also for the heavy nuclei.
The relative deviations of the Coulomb exchange energies in the
calculations with phenomenological effective charge factors are less than
$1\%$ for traditional semi-magic Ca, Ni, Sn, and Pb isotopes from proton
drip line to neutron drip line. Furthermore, one found that the BED of
mirror nuclei are sensitive to the Coulomb exchange term, and the
discrepancy between the BED of mirror nuclei calculated by the RHB
approach with the RH parameter sets and those calculated by the RHFB
approach with PKA1 can be remarkably reduced by using the phenomenological
effective charge factors.

%-------------------------------------------------------------------------------------------------------
%\section{Acknowledgements}
We thank S. Q. Zhang, H. Z. Liang, and P. W. Zhao for stimulating
discussions. This work was partly supported by the National Natural
Science Foundation of China under Grant Nos. 11205004, 11175001, and
11075066, the 211 Project of Anhui University under Grant No.
02303319-33190135, the Program for New Century Excellent Talents in
University of China under Grant No. NCET-05-0558, and the Talent
Foundation of High Education of Anhui Province for Outstanding Youth under
Grant No. 2011SQRL014.
%-------------------------------------------------------------------------------------------------------

%-------------------------------------------------------------------------------------------------------

%-------------------------------------------------------------------------------------------------------

%-------------------------------------------------------------------------------------------------------
\end{CJK*}
\end{document}